\documentstyle[amssymb,preprint,eqsecnum,prb,aps,epsfig]{revtex}
\tightenlines

\begin{document}
\title{Penetration Depth and Anisotropy in MgB$_{2}$}
\author{X. H. Chen, Y. Y. Xue, R. L. Meng and C. W. Chu$^{1}$}
\address{Department of Physics and Texas Center for Superconductivity,\\University of
Houston, Houston, TX 77204-5932\\$^{1}$ also at Lawrence Berkeley National 
Laboratory, 1 Cyclotron Road,\\Berkeley, CA 94720}
\date{\today}
\maketitle
\pacs{74.25Ha, 74.25Bt}

\begin{abstract}
The penetration depth $\lambda $ of MgB$_{2}$ was deduced from both the {\it %
ac} susceptibility $\chi $ and the magnetization $M(H)$ of sorted powders.
The good agreement between the two sets of data without geometric correction
for the grain orientation suggests that MgB$_{2}$ is an isotropic
superconductor.
\end{abstract}


Great interest has been raised recently by the discovery\cite{nag} of MgB$%
_{2}$ with T$_{c}$ about $40~$K. In its normal state, the compound appears
to be a metal with a low {\it dc} resistivity,\cite{can} Hall coefficient,
\cite{kan} and thermoelectric power\cite{lor} dominated by hole carriers.
The estimated long mean-free path\cite{can} implies that the electrical
transport could well be isotropic in spite of its layer-like crystalline
structure, in agreement with the band-structure calculations.\cite{kor}
Below a transition temperature of T$_{c}$ $\approx $ 40~K, the compound
seems to be a phonon-mediated BCS superconductor, as suggested by the large
isotopic effect\cite{bud} and the large negative pressure effect\cite{lor}
on T$_{c}$. The upper critical field, H$_{c2}$(T), has been
directly measured above 20 K with a linear (or even an upward curvature) $T$%
-dependence.\cite{fin} The lower critical field was either indirectly
estimated based on the thermodynamics field H$_{c}$(T) above 34~K or
calculated based on the non-linearity in $M(H)$ at low temperatures.\cite
{fin,tak} The coherence length $\xi _{\text{o}}$(0~K), penetration depth $%
\lambda $(0~K), and Ginzburg-Landau parameter $\kappa $ were consequentially
estimated to be 5.2~nm, 125-140~nm, and 26, respectively. All these
calculations, however, were based on the assumption that MgB$_{2}$ can be
treated as isotropic. Otherwise, a geometric factor up to 2 would be needed
to correct the random grain-orientation and to convert $M_{r}$ to $H_{c}$.

To estimate the anisotropy, $\lambda $ of MgB$_{2}$ was directly measured
using both $ac$ susceptibility and the non-linearity in the $M(H)$ of powder
samples. Although the two procedures involve the anisotropy of $\lambda $ in
very different ways, the $\lambda $(5~K)'s deduced from these procedures are
in good agreement without geometric corrections for superconducting anisotropy. The
results, therefore, suggest that the anisotropy of MgB$_{2}$ is very small.

Ceramic MgB$_{2}$ samples were prepared using the solid-state reaction
method.\cite{bud} Small Mg chips (99.8\% pure) and B powder (99.7\% pure)
with a stoichiometry of Mg:B = 1:2 were sealed inside a Ta tube under an Ar
atomosphere. The sealed Ta ampoule was then enclosed in a quartz tube. The
assembly was heated slowly up to 950~$^{\text{o}}C$ and was kept at this
temperature for 2 hours, followed by furnace cooling. The structure was
detemined by X-ray powder diffraction (XRD) using a Rigaku DMAX-IIIB
diffractometer. Powder samples were prepared by sorting the pulverized
powder using either sieves or the method of descending speed of the
particles in acetone. No grain alignment was attempted. The grain morphology
as well as the particle sizes of the powders were measured using a JEOL JSM
6400 scanning electron microscope (SEM). Magnetizations were measured in a
Quantum-Design 5 T SQUID magnetometer with an {\it ac} attachement. The {\it %
ac} susceptibility was measured under a fixed frequency {\it f} = 413.1~Hz
and an amplitude of 3~Oe.

The XRD pattern can be indexed as a hexagonal cell with lattice parameters
{\it a} = 3.08~\AA\ and {\it c} = 3.52~\AA . A sharp superconducting
transition was observed in both resistivity and $ac$ susceptibility with T$%
_{c}$ $\approx $ 38~K.\cite{lor}

The transition temperature in different fields, $T_{c}(H)$, was determined
from both the {\it dc} magnetization and the {\it ac} suscepitibility
measurements, which should be equivalent to a local resistivity measurement,
with a {\it dc} bias of 0 - 5~T. The transition width in the {\it ac}
suscepitibility is only slightly broadened under fields, {\it i.e.} from $%
\approx $ 1~K at the bias field of $H_{dc}$ = 0~T to $\approx $ 5~K at 5~T
(Inset, Fig. 1). We attribute this to the flux movement under fields and
take the onset temperature as $T_{c}(H)$ (Fig. 1). The data from the two
methods are reasonably consistent with a slope $dH_{c2}/dT$ $\approx $ 0.4$%
\pm 0.05$~T/K near 38~K, in agreement with the results obtained by Finnemore
{\it et al}.\cite{fin}

The lower critical field was measured in powder samples with a particle size
\mbox{$<$}%
2~$\mu $m to avoid the complications caused by intergrain coupling.\cite
{note} No systematic variation between the deduced $\lambda $(T) and the
particle sizes was observed. This demonstrates that the grain-boundary
effect in the sample is small, a fact that is also supported by direct SEM
observation of the powders.

It is known that the $ac$ susceptibility $\chi $ of a superconducting sphere
of diameter $d$ is $\Phi (d,\lambda )$ = $-3/(8\pi )[1-6(\lambda
/d)coth(d/2\lambda )+12(\lambda /d)^{2}]$, which reduces to $\propto $ $%
(d/\lambda )^{2}$ when $d<2\lambda $. Methods have been previously developed
to deduce $\lambda $ from the $\chi $ of magnetically aligned powders by
solving the equation $\chi $ = $\Phi (d,\lambda )$ $\approx $ -0.002$%
(d/\lambda )^{2}$ in the large-$\lambda $ limit.\cite{por} However, the $%
\lambda $ so deduced will be sensitive to the uncertainty of $\chi $ (due to
either the demagnetization factor or the superconducting volume fraction) if
$d$
\mbox{$>$}%
\mbox{$>$}%
\ 2$\lambda $ as in unsorted powders, which may include particles as large
as 3 $\mu $m.\cite{por} In fact, the calculated $\partial \ln \lambda
/\partial \ln \chi $ varies with $d/\lambda $ only moderately below $%
d/\lambda $ = 4, {\it i.e} from -0.5 to -0.7, but changes rapidly for
larger $d/\lambda $. For example, $\partial \ln \lambda /\partial \ln \chi $
is -3 at $d/\lambda $ = 20, and a 30\% uncertainty of $\chi $ will lead to a
$\lambda $ anywhere between 0 and 10$d$. A $d/\lambda $
\mbox{$<$}%
5 is needed to obtain a 20\% accuracy with an estimated 30\% uncertainty in $%
\chi $. The technique can be improved by using sorted powders, which have a
smaller $d$ and a narrower size-distribution.\cite{xue} The powder obtained
from pulverizing ceramic was thoughly mixed with acetone in a 10 ml beaker.
The particles was then sorted according to the time needed for them to reach
the bottom of the beaker. The sample discussed here was collection of
particles deposited between 1-2 hr. Our SEM observation suggested that 99\%
or more particles having a size between 0.1 and 2 $\mu $m (Fig. 2). It
should also be noted that the proposed method of calculating the effective
grain-size {\it d} = $\sqrt{(\sum \text{{\it d}}_{i}^{5})/(\sum \text{{\it d}%
}_{j}^{3})}$ (where {\it d}$_{i}$ is the diameter of individual grains) may
also be questionable if $d$
\mbox{$>$}%
$2\lambda $.\cite{por} As will be shown below, a 30\% error may be cuased by
the approaximation alone. A regression procedure, therefore, was adopted. A $%
\lambda _{raw}$ corresponding to the {\it d}$_{raw}$ = $\sqrt{(\sum \text{%
{\it d}}_{i}^{5})/(\sum \text{{\it d}}_{j}^{3})}$ was used as the initial
value. The $d$ was then refined regressively as $[\sum \Phi (\lambda ,d_{i})$%
{\it d}$_{i}^{4}]/[\sum \Phi (\lambda ,d_{i})${\it d}$_{i}^{3}]$. The
convergence is very fast. It should be noted that the effective $d$ depends
on $\lambda $, {\it i.e.} the correction varies with T and may change the
T-dependence of $\lambda $. Our tests on YBa$_{2}$Cu$_{3}$O$_{7-\delta }$
powders demonstrated that the uncertainty of the $\lambda $ so deduced is
within 10-20\% of the published data if $d/\lambda $ is 3 or smaller.\cite
{xue}

The SEM photo of a powder sample is shown in Fig. 2. Its $d$-distribution is
shown in the inset of Fig. 3. A {\it d} $\approx $ 0.88~$\mu $m was obtained
with {\it d}$_{raw}$ = 1.23~$\mu $m. The superfluid density 1/$\lambda
^{2}(T)$ was calculated assuming an isotropic superconductivity, {\it i.e. }%
without corrections for the random grain orientation (Fig. 3). The 1/$%
\lambda ^{2}(T)$ observed roughly follows a T-dependency of [1-($T$/39.4)$%
^{2.7}$]. It should be noted that a deviated from the fit is clear below 10
K. We are hesitate, however, to draw any conclusion about the deviation
since weak-links can not be conclusively excluded at this stage. The
extrapolated $\lambda $(0) $\approx $ 180~nm is slightly longer than that
found by Finnemore {\it et al}. \cite{fin}

The lower critical field $H_{c1}$ was also deduced as the field, where the
linear $M-H$ correlation begins to be violated, from the magnetizations of
the same powder sample in an $H$-increase branch at 5~K (Inset, Fig. 4).
Several technical difficulties in this method have been previously
discussed: for instance, the intergrain coupling that may cause nonlinearity
far below the intragrain $H_{C1}$; the surface pinning that can make the $%
H_{C1}$ value observed higher; sharp local edges,{\it \ i.e. }strong local
demagnetizing fields, that can lead to a lower one; and the experimental
resolution of the nonlinearity. Several precautions have been taken. The
powder sample used here has a particle size far smaller than the average
grain size, which should eliminate the effect of the intergrain coupling, as
suggested by the smooth and flat $\chi $ observed below T$_{c}$. To improve
the sensitivity of nonlinearity, the $M(H$, 5~K) below 50~Oe was fit as a
linear function of $H$ using a standard least-square procedure. The
deviation $\Delta $M ($\approx $ 0.002~emu/cm$^{3}$ below 50~Oe) from the
linear fit is comparable with the experimental uncertainties in both $M$ and
$H$, demonstrating the negligible effect of the residual granularity. The
difference between the data and the extrapolated linear fit above 50~Oe was
then calculated. The uncertainty associated with the linear fitting was
marked as dashed lines in Fig. 4. To further avoid the interferences from
the sharp edges of the particles, the deviation at large fields was
empirically fit as $a\cdot (H-H_{\text{o}})^{1.8}$ with both $a$ and $H_{%
\text{o}}$ as the free parameters (Fig. 4). We justify the fit by pointing
out that the magnetization of a superconductor partially penetrated by
external fields will vary as the square of the thickness penetrated, {\it i.e%
} $H-H_{\text{o}}$ in the Bean model. It should be pointed out that the value
of $H_{\text{o}}$ is not very sensitive to the index chosen. A linear fit
below 200 Oe leads to only 15\% change. Surface pinning is usually
negligible in randomly shaped grains, and typically can only make the H$_{c1}
$ observed smaller. Following the procedure, an H$_{c1}$ = $H_{\text{o}%
}/(1-g)\approx $ 130~Oe was obtained, where $g$ = 1/3 is the demagnetization
factor of a sphere. This value consequently leads to a $\lambda $ of 203~nm,
in good agreement with that from the{\it \ ac} $\chi $ within the
uncertainty of the techniques. No corrections have been made to consider
random grain orientations.

To further verify the result, the {\it ac} susceptibility of the same powder
sample was measured at 5~K with a {\it dc} bias between 0 and 200~Oe (Fig.
5). A change of the slope was observed around H$_{\text{o}}$ $\approx $
110~Oe, and the H$_{c1}$ was estimated as $\approx $ 160~Oe. Similar
measurements have been done in several different samples and the results
appear to be independent of the particle sizes.

The deduced $\lambda $ is slightly longer than the 140~nm\cite{fin} and the $%
\approx $ 130~nm\cite{tak} previously reported. The exact reason for the
disagreement is not clear to us at this moment. However, the $\lambda $
measured here using three different methods on the same sample are
self-consistent within the estimated experimental uncertainty of $\pm $20\%.

It is interesting to note the agreement between the H$_{c1}$ from $\chi $
and $M(H)$. In a highly anisotropic layered superconductor, cuprates for
example, the observed $\chi $ will only come from the supercurrents in the
layers. The $\chi $ observed, therefore, should be assumed to be $\int \cos
^{2}\theta \sin \theta d\theta /\int \sin \theta d\theta $ $\approx $ 1/3 of
the $-3/(8\pi )[1-6(\lambda /d)coth(d/2\lambda )+12(\lambda /d)^{2}]$. The
deduced $\lambda $ will be 1.7 times longer if no geometric correction has
been made. In general, the $1$/$\lambda ^{2}$ deduced from non-grain-aligned
powder should be 1/3$\lambda _{ab}^{2}$+2/3$\lambda _{c}^{2}$ = (1/3+2/3$%
\gamma )/\lambda _{ab}^{2}$ in layered superconductors, where $\lambda _{ab}$%
, $\lambda _{c}$, and $\gamma $ are the penetration depths in and out of the
layers, and the anisotropy, respectively. The H$_{c1}$ deduced from the
nonlinearity of $M-H$, on the other hand, can be even larger since the
effective field perpendicular to the local layers is only a fraction, {\it %
i.e}. cos$\theta $, of the external field. The good agreement observed here,
therefore, strongly suggests that MgB$_{2}$ is an isotropic superconductor.
The estimated $\gamma $ should be smaller than 1.5 assuming a experimental
uncertainty of $\pm $20\% in our $\lambda $-calculation. This $\gamma $ is
far smaller than that of 10-1000 observed in various cuprates, and should be
regarded as essentially isotropic.

In above data analysis, a spherical shape was assumed. The relative change
of the demagnetization factor is $\Delta $r/3r in a slightly-deformed
ellipsoid with radii r+$\Delta $r, r+$\Delta $r/2, and r+$\Delta $r/2. A
simple calculation shows that the correction of the H$_{c1}$ will be -0.25($%
\Delta $r/r)$^{2}$ in the {\it ac} $\chi $ method,\cite{note1} but $\Delta $%
r/3r in the nonlinearity method. An average length ratio (r+$\Delta $r)/(r-$%
\Delta $r) between 0.5 to 2 ({\it i.e.} $\mid \Delta $r/3r$\mid =1/3$),
therefore, may not significantly changes above conclusion. The condition
seem to be satisfied (Fig. 2).

This conclusion is in agreement with the band structure calculation, the
extremely long mean-free path, the long coherence length, and the small
grain-boundary effect on the supercurrents reported.

In summary, the penetration depth $\lambda $(T) of MgB$_{2}$ was deduced
from both the $ac$ susceptibility $\chi $ of powders and the nonlinearity of
the $M-H$ in the $H$-increase branch. The good agreement between the two
methods suggests that MgB$_{2}$ is an isotropic superconductor.

This work is supported in part by the NSF, the T. L. L. Temple Foundation,
the John and Rebecca Moores Endowment and the State of Texas through TCSUH,
and at LBNL by DOE.

\begin{figure}[t]
\caption{H$_{c2}$ of a MgB$_{2}$ ceramic sample. $\blacktriangledown $: from
the $dc$ magnetization; $\bullet $: from the $ac$ susceptibility with a $dc$
bias $H$. Inset: the $ac$ susceptibility at H = $\blacksquare $: 0~T; $%
\blacktriangledown $: 2.5~T, and $\bullet $: 5~T. }
\label{Fig. 1}
\end{figure}
\begin{figure}[t]
\caption{SEM photo of the powder sample}
\label{Fig. 2}
\end{figure}

\begin{figure}[t]
\caption{1/$\protect\lambda ^{2}$($T$) of a MgB$_{2}$ powder sample. $%
\bigcirc $: data; solid line: fit as $\propto $ [1-($T$/39.4)$^{2.7}$].
Inset: the particle-size distribution of the powder.}
\label{Fig. 3}
\end{figure}

\begin{figure}[t]
\caption{The deviation, $\Delta M$, from the linearly extrapolation at 5 K. $%
\bigcirc $: data; dashed lines: the uncertainty bands of the linear fit;
solid line: a fit of $(H-H_{\text{o}})^{1.8}$. Inset: $M(H)$ at 5 K.}
\label{Fig. 4}
\end{figure}

\begin{figure}[t]
\caption{The $ac$ susceptibility with a $dc$ bias $H$ at 5 K.}
\label{Fig. 5}
\end{figure}


\begin{references}
\bibitem{note1}  The first-order correction should be zero due to the random
orientations of the particles.
\end{references}
\end{document}